\documentclass[12pt]{article}
\usepackage{a4wide,graphicx}

\newcommand{\be}{\begin{equation}}
\newcommand{\ee}{\end{equation}}
\newcommand{\ba}{\begin{eqnarray}}
\newcommand{\ea}{\end{eqnarray}}

\begin{document}
\begin{titlepage}
\mbox{} 
\vspace*{2.5\fill} 
{\Large\bf 
\begin{center}
%
Influence of nucleonic motion in Relativistic Fermi Gas 
inclusive responses
%
\end{center}
} 

\vspace{1\fill} 

\begin{center}
{\large 
L. Alvarez-Ruso$^{1}$,
M.B. Barbaro$^{1}$, 
T.W. Donnelly$^{2}$ and 
A. Molinari$^{1}$
}
\end{center}

\begin{small}
\begin{center}
$^{1}${\sl 
Dipartimento di Fisica Teorica,
Universit\`a di Torino and
INFN, Sezione di Torino \\
Via P. Giuria 1, 10125 Torino, ITALY 
}\\[2mm]
$^{2}${\sl 
Center for Theoretical Physics, \\
Laboratory for Nuclear Science and Department of Physics\\
Massachusetts Institute of Technology,
Cambridge, MA 02139, USA 
}
\end{center}
\end{small}

\kern 1. cm \hrule \kern 3mm 

\begin{small}\noindent
{\bf Abstract} 
\vspace{3mm} 

Impulsive hadronic descriptions of electroweak processes in
nuclei involve two distinctly different elements: one stems 
from the nuclear many-body physics --- the medium --- which
is rather similar for the various inclusive response functions,
and the other embodies the responses of the hadrons themselves
to the electroweak probe and varies with the channel selected.
In this letter we investigate within the context of the
relativistic Fermi gas in both the quasi-elastic and $N\to\Delta$
regimes the interplay between these two elements. Specifically,
we focus on expansions in the one small parameter in the problem,
namely, the momentum of a nucleon in the initial wave function
compared with the hadronic scale, the nucleon mass. Both 
parity-conserving and -violating inclusive responses are
studied and the interplay between longitudinal ($L$) and transverse
($T$ and $T'$) contributions is highlighted.

\vspace{2cm}

\kern 2mm 


\end{small}

\end{titlepage}

In this letter we explore the influence of 
relativistic kinematics and of the medium 
on inclusive electron-nucleus scattering in the context of
the Relativistic Fermi Gas (RFG). In spite of its simplicity 
the RFG is covariant, a requirement that is clearly relevant for
analyses of  modern
medium- to high-energy nuclear electron scattering experiments.
In past work compact, transparent expressions have been obtained
for the longitudinal $(L)$ and transverse $(T)$ responses in 
parity-conserving
(PC) scattering and the $L$, $T$ and axial $(T')$ response functions in
parity-violating (PV) scattering. Both the quasi-elastic (QE)
and $\Delta$ regions have been studied \cite{Don92,Ama99,Amo00}. 
In all cases the responses factorize as follows
\ba
R^{L,T}(q,\omega) &=& R_0(q,\omega) U^{L,T}(q,\omega) 
\label{respPC}
\\
\tilde R^{L,T,T'}(q,\omega) &=& R_0(q,\omega) \tilde U^{L,T,T'}(q,\omega) 
\label{respPV}
\ea
for the PC and PV sectors, respectively, providing a unified 
approach for treatments of electroweak interactions at high inelasticity.
In (\ref{respPC}) and (\ref{respPV}) the $\Delta$ is viewed as a stable 
particle,
but later we shall model its decay by including its width. 

Defining an overall constant $C\equiv 3{\cal N}/(4\kappa m_N\eta_F^3)$, 
here we have a universal factor (the same for all the responses we consider)
\be
R_0(q,\omega)=C \xi_F (1-\psi^2)\,,
\label{R0}
\ee
which scales in the region where the RFG is expected to be a
reasonable model, specifically at $q>2 p_F$ where Pauli blocking
effects are absent~\cite{Alb88,Bar98,Don99}. That is, it depends 
only upon a single variable $\psi$, whose square is given by
\be
\psi^2=\frac{\epsilon_0-1}{\xi_F}=
\frac{1}{\xi_F}\left(\kappa\sqrt{\frac{1}{\tau}+\rho^2}-\lambda\rho-1\right)
\ ,
\label{psi}
\ee
linked to the
minimum energy $\epsilon_0$ a nucleon can have in the RFG to participate 
in the response of the system to an external electroweak field.
In contrast to $R_0$, 
the functions $U^{L,T,T'}$ depend upon the specific electron-nucleon 
process of interest.

In the above ${\cal N}$ is the number of nucleons 
\footnote{Actually in the nucleonic sector one should add, with the 
appropriate form factors, the responses with ${\cal N}=Z$ and ${\cal N}=N$
separately; in the $\Delta$ sector one should simply set ${\cal N}=A$.},
$\eta_F=\sqrt{\xi_F(\xi_F+2)}=p_F/m_N$ is the dimensionless Fermi momentum 
and the dimensionless four-momenta
\be
\eta^\mu=(\epsilon,\vec\eta)=
\left(\sqrt{1+p^2/m_N^2},\ \frac{\vec p}{m_N}\right)\ ,
\ \ \ \ 
\kappa^\mu=(\lambda,\vec\kappa)=
\left(\frac{\omega}{2m_N},\frac{\vec q}{2 m_N}\right)
\ee
have been introduced for the initial nucleon and the exchanged boson, 
respectively, with $\tau=\kappa^2-\lambda^2$.
Moreover, $\rho$ is an inelasticity parameter given by \cite{Ama99} 
\be
\rho = 1+\frac{1}{4 \tau}(\mu^2-1) \ ,
\label{rho}
\ee
with $\mu\equiv m_\Delta/m_N$, which reduces to unity in the nucleonic 
quasi-elastic sector.

The response functions (\ref{respPC}) and (\ref{respPV})
are obtained as specific components of the nuclear tensor
\be
W_{\mu \nu} = \frac{3{\cal N}}{4\pi m_N\eta_F^3} \int
\frac{d \vec\eta}{\epsilon \epsilon'} \, 
\delta(2 \lambda - \epsilon' + \epsilon)
\, \theta(\eta_F - |\vec\eta|) f_{\mu \nu}(\vec\eta)\ ,
\label{Wmunu}
\ee
which incorporates the single-nucleon responses, expressed 
via the single-nucleon tensor $f_{\mu\nu}$, over the allowed energy range.
In particular, one has
\ba
R^L &=& W_{00} = 
C\int_{\epsilon_0}^{\epsilon_F} f_{00}(\epsilon,\theta_0) d\epsilon
\label{RL}
\\
R^T &=& W_{11}+W_{22} =
C\int_{\epsilon_0}^{\epsilon_F} 
[f_{11}(\epsilon,\theta_0)+f_{22}(\epsilon,\theta_0)] d\epsilon
\label{RT}
\\
\tilde R^L &=& \tilde W_{00} = 
C\int_{\epsilon_0}^{\epsilon_F} \tilde f_{00}(\epsilon,\theta_0) d\epsilon
\label{RLPV}
\\
\tilde R^T &=& \tilde W_{11}+\tilde W_{22} =
C\int_{\epsilon_0}^{\epsilon_F} 
[\tilde f_{11}(\epsilon,\theta_0)+\tilde f_{22}(\epsilon,\theta_0)] d\epsilon
\label{RTPV}
\\
\tilde R^{T'} &=& -i \tilde W_{12} = -i C\int_{\epsilon_0}^{\epsilon_F} 
\tilde f_{12}(\epsilon,\theta_0) d\epsilon\ ,
\label{RTP}
\ea
where
\be 
\cos\theta_0=\frac{\lambda\epsilon-\tau\rho}{\kappa\eta}
\label{cos}
\ee 
and
\be
f_{\mu\nu} (\epsilon,\theta_0)=
 -w_1(\tau)\left(g_{\mu\nu}+\frac{\kappa_\mu\kappa_\nu}{\tau}\right)
             +w_2(\tau) V_\mu V_\nu
             -2iw_3(\tau)\epsilon_{\mu\nu\rho\sigma}\kappa^\rho V^\sigma \ ,
\label{fmunu}
\ee
with
$V_\mu = 
\eta_\mu+\kappa_\mu \rho$. Note that (\ref{cos}) follows from energy
conservation and the lower limit of integration is defined by (\ref{psi}).
The single-nucleon tensor (\ref{fmunu}) is valid for both the 
quasi-elastic process and the
$N\to\Delta$ transition, but of course the $w_i$'s are different 
in the two cases. 
In the QE region these are \cite{Don92} 
\ba
w_{1,a}(\tau) &=& \tau G_{M,a}^2(\tau)
\\
w_{2,a}(\tau) &=& \frac{1}{1+\tau} 
\left[G_{E,a}^2(\tau)+\tau G_{M,a}^2(\tau)\right]
\ea
in the PC sector and
\ba
\tilde w_{1,a}(\tau) &=& \tau G_{M,a}(\tau)\tilde G_{M,a}(\tau)
\\
\tilde w_{2,a}(\tau) &=& \frac{1}{1+\tau}
\left[G_{E,a}(\tau)\tilde G_{E,a}(\tau) +
\tau G_{M,a}(\tau)\tilde G_{M,a}(\tau)\right]
\\
\tilde w_{3,a}(\tau) &=& G_{M,a}(\tau) \tilde G_{A,a}(\tau)
\ea
in the PV one, the index $a$ referring
to protons or neutrons. 

For the $\Delta$ sector one has instead
\ba
w_{1,\Delta}(\tau) &=& \frac{1}{16} \left[4\tau+(\mu-1)^2\right]
(\mu+1)^2 
\left[G_{M,\Delta}^2(\tau)+3 G_{E,\Delta}^2(\tau)\right]\\
w_{2,\Delta}(\tau) &=& \frac{1}{16} \frac{4\tau+(\mu-1)^2}{1+\tau\rho^2} 
(\mu+1)^2
\left[G_{M,\Delta}^2(\tau)+3 G_{E,\Delta}^2(\tau)+
\frac{4\tau}{\mu^2}G_{C,\Delta}^2(\tau)\right] \nonumber\\
&&
\ea
in the PC sector \cite{Ama99} and
\ba
\tilde w_{1,\Delta}(\tau) &=& \frac{1}{16} \left[4\tau+(\mu-1)^2\right]
(\mu+1)^2  \nonumber\\
&&\times \left[G_{M,\Delta}(\tau) \tilde G_{M,\Delta}(\tau) +
3 G_{E,\Delta}(\tau) \tilde G_{E,\Delta}(\tau) \right] \\
\tilde w_{2,\Delta}(\tau) &=&  
\frac{1}{16} \frac{4\tau+(\mu-1)^2}{1+\tau\rho^2} 
(\mu+1)^2 \nonumber\\
&& \times \left[G_{M,\Delta}(\tau) \tilde G_{M,\Delta}(\tau) +
3 G_{E,\Delta}(\tau)  \tilde G_{E,\Delta}(\tau)  +
\frac{4\tau}{\mu^2} G_{C,\Delta}(\tau) \tilde G_{C,\Delta}(\tau) \right]
\nonumber\\
&& \\
\tilde w_{3,\Delta}(\tau) &=& \frac{1}{4}(\mu^2-1)
[3  G_{E,\Delta}(\tau) \tilde G_{M,\Delta}^A(\tau) 
+ G_{M,\Delta}(\tau) \tilde G_{E,\Delta}^A(\tau) ]
\ea
in the PV one \cite{Amo00}. The parametrizations of the elastic 
form factors are given in \cite{Don92}, while those for the
$N\to\Delta$ transitions are the following:
\ba
G_{E,\Delta}(\tau) &=& -0.03\, f(\tau)
\\
G_{M,\Delta}(\tau) &=&  2.97\, f(\tau)
\\
G_{C,\Delta}(\tau) &=& 0
\\
\tilde G_{E,\Delta}(\tau) &=& \beta^{T=1}_V G_{E,\Delta}(\tau)
\\
\tilde G_{M,\Delta}(\tau) &=& \beta^{T=1}_V G_{M,\Delta}(\tau)
\\
\tilde G_{C,\Delta}(\tau) &=& \beta^{T=1}_V G_{C,\Delta}(\tau)
\\
\tilde G^A_{E,\Delta}(\tau) &=& 2.22\, G_D^A(\tau)
\\
\tilde G^A_{M,\Delta}(\tau) &=& 0\ ,
\ea
where $f(\tau) = G_{E,p}(\tau) (1+\lambda_\Delta \tau)^{-1/2}, $
with $\lambda_\Delta=1.0$ and
$\beta_V^{T=1}=1 - 2\sin^2{\theta_W} = 0.55$.

The energy integrations in (\ref{RL}-\ref{RTP}) are easily performed
and yield the well-known results \cite{Don92,Ama98,Amo00}
\ba
&&U^L(\kappa,\tau) = \frac{\kappa^2}{\tau} 
\left[(1+\tau\rho^2) w_2(\tau)-w_1(\tau) +w_2(\tau){\cal D}^L(\kappa,\tau)
\right] 
\label{UL}\\
&&U^T(\kappa,\tau) = 2 w_1(\tau) +w_2(\tau){\cal D}^T(\kappa,\tau) 
\label{UT}\\
&&\tilde U^L(\kappa,\tau) = \frac{\kappa^2}{\tau} 
\left[(1+\tau\rho^2) \tilde w_2(\tau)-\tilde w_1(\tau) +
\tilde w_2(\tau){\cal D}^L(\kappa,\tau)\right] 
\label{ULPV}\\
&&\tilde U^T(\kappa,\tau) = 2 \tilde w_1(\tau) +\tilde w_2(\tau)
{\cal D}^T (\kappa,\tau)\label{UTPV}\\
&&\tilde U^{T'}(\kappa,\tau) = 2 \sqrt{\tau(\tau\rho^2+1)}
 \tilde w_3(\tau)\left[1+{\cal D}^{T'}(\kappa,\tau)\right] 
\label{UTP}\ ,
\ea
where the functions ${\cal D}^{L,T,T'}$ are such as to vanish in the 
limit $\xi_F\sim \eta_F^2\to 0$. Note that expressions such as
(\ref{UL}) contain factors of the type $\kappa^2/\tau$. If the $\kappa^2$
in these factors were also expanded using, for example, $\tau$, $\eta$
and $\rho$ as independent variables, then terms linear in $\eta$
(and hence $\eta_F$) would occur. However, when written in the form
given here using both $\kappa$ and $\tau$, more compact expressions
result and the corrections --- the terms containing the ${\cal D}$'s
--- incur only small contributions of order $\eta_F^2$, namely, only
quadratic in the one small parameter at our disposal.

We now explore the impact of the nucleonic motion on the inclusive
responses. We begin by expanding the tensor $f_{\mu\nu}$ in the 
dimensionless nucleon 
momentum $\eta$ up to linear order, in the same spirit as in
\cite{Jes98,Ama98,Ama99} where expansions have been performed
for the spinor matrix elements of the nucleon electromagnetic single-nucleon
and meson-exchange currents. 
Importantly, in the framework of this expansion one has that
\be
\kappa\eta\cos\theta_0\simeq \lambda-\tau\rho \ ,
\ee
if terms higher than linear order are neglected.
Since the $w_i$'s only depend upon the external variable $\tau$,
and the leading contribution in the nucleonic momentum expansion is
quadratic for ${\cal D}^{L,T,T'}$, as we shall see, then the following
expressions for the $U$ functions are obtained from (\ref{UL})-(\ref{UTP})
\ba
&&U^L(\kappa,\tau)\simeq 
\left[ (1 + \tau\rho^2) + 2\rho(\lambda - \tau\rho) \right] 
\left[(1+\tau\rho^2) w_2(\tau)-w_1(\tau) \right] 
\label{ULE}
\\
&&U^T(\kappa,\tau) \simeq 2 w_1(\tau) 
\label{UTE}
\\
&&\tilde U^L(\kappa,\tau)\simeq \left[ (1 + \tau\rho^2) +
 2\rho(\lambda - \tau\rho) \right] 
\left[(1+\tau\rho^2) \tilde w_2(\tau)-\tilde w_1(\tau) \right] 
\\
&&\tilde U^T (\kappa,\tau)\simeq 2 \tilde w_1(\tau) \\
&&\tilde U^{T'}(\kappa,\tau) \simeq 2 \sqrt{\tau(1+\tau\rho^2)}
 \tilde w_3(\tau) \ .
\label{UTPE} 
\ea

\begin{figure}[h!]
\begin{center}
\includegraphics[height=\textwidth, angle=-90]{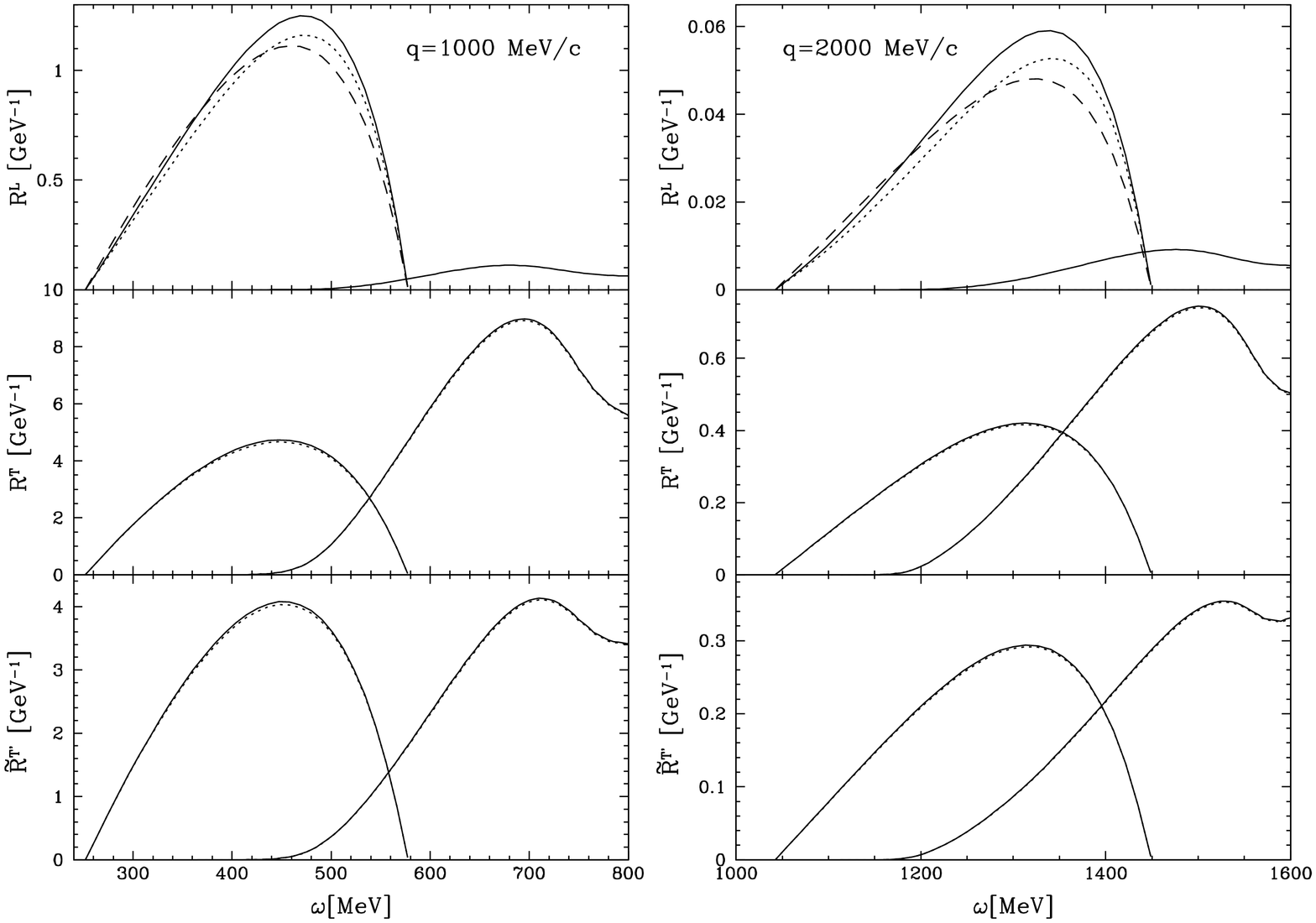}
\caption{Electromagnetic and PV axial responses for $^{12}C$ ($p_F =225~MeV/c$).
Every plot shows both the QE (left peak) and $\Delta$ (right peak) regions. 
The solid lines correspond to the exact responses from
(\ref{UL}, \ref{UT}, \ref{UTP}); the dotted lines show the result of the
expansion in $\eta$ up to linear terms ((\ref{ULE}, \ref{UTE},
\ref{UTPE})); the dashed ones keep only the leading order in  $\eta$ (which is
different from the linear one only for the longitudinal response).}       
\label{figup}
\end{center}
\end{figure}

In the above, only $U^L$ and $\tilde U^L$ turn out to be 
affected by the terms linear in the struck nucleon's 
momentum $\eta$ through the expression $\lambda-\tau\rho$,
whereas the transverse and axial functions are not.
Higher-order terms in the $\eta$ expansion will modify $U^T$, $\tilde U^T$ 
and $\tilde U^{T'}$ only as far as 
${\cal D}^T$ and ${\cal D}^{T'}$ are concerned, 
in contrast with the longitudinal case where they 
affect the whole of $U^L$ and $\tilde U^L$.
In order to illustrate these items we show in Fig.~\ref{figup} the
electromagnetic and PV axial responses for both the QE and $N\to\Delta$ processes.
The responses in the $\Delta$ region take into account the finite width of the
resonance by the standard substitution 
\be
R(q,\omega) \rightarrow \int_{W_{min}}^{W_{max}} \frac{1}{2 \pi}
\frac{\Gamma (W)}{(W - m_\Delta)^2 + \Gamma^2 (W)/4} R(q,\omega, W) dW \,, 
\ee
where $R(q,\omega,W)$ is the response for a stable $\Delta$ of mass $W$;
the invariant mass of the resonance $W$ ranges from $W_{min}=m_N + m_\pi$ to
$W_{max}=\sqrt{(E_F + \omega)^2 - (q - p_F)^2}$ and $\Gamma (W)$ is the usual
P-wave $\Delta$ width \cite{OTW}. We plot the exact RFG results given by
(\ref{UL}, \ref{UT}, \ref{UTP}) and those obtained from our expansion in
$\eta$ of (\ref{ULE}, \ref{UTE}, \ref{UTPE}). In the case of the
transversal and axial responses, the approximate result can hardly be
distinguished from the exact one, showing that they are practically 
unaffected by the higher-order terms in $\eta$ 
(i.e., by ${\cal D}^{T(T')}$). The
situation is different for the longitudinal response, as can be appreciated
for the QE peak. Only here there is a contribution from linear terms, which
improves the result of the leading-order
expression (given by (\ref{ULE}) without $2 \rho (\lambda - \tau \rho)$),
but it is still insufficient to reproduce the exact result. On the other hand,
both leading and next-to-leading
$N\to\Delta$ longitudinal responses vanish in the absence of
${\cal D}^{L}$. This is a consequence of having $G_{C,\Delta}(\tau)=0$, which
implies $(1+\tau\rho^2) w_{2,\Delta} (\tau)-w_{1, \Delta}(\tau)=0$.
Therefore, in the absence of the Coulomb form factor, the
longitudinal $N\to\Delta$ response is proportional to ${\cal D}^{L}$. A
non-vanishing $G_{C,\Delta}(\tau)$, even if small, avoids the
cancellation and produces a contribution to $U^L$ in the $\Delta$ region that
might be relevant at large momentum transfer~\cite{Ama99}.           
  
Interestingly, the parts of $U^L$ and $\tilde U^L$ not proportional to 
${\cal D}^L$ arise entirely
from the longitudinal component of the nucleonic momentum
\be
\eta_L \simeq \frac{1}{\kappa} (\lambda-\tau\rho)\ .
\ee
On the other hand the transverse component of $\vec\eta$, namely
\be
\eta_T = 
\sqrt{\frac{\tau}{\kappa^2}(\epsilon+\lambda\rho)^2-1-\tau\rho^2} \ ,
\ee
only contributes to ${\cal D}^L$. 
Actually, the longitudinal and transverse motion of a nucleon inside the RFG
are not disconnected, but turn out to be linked by the
kinematical constraint
\be
\eta_L = \lambda\sqrt{\frac{1}{\tau}(1+\tau\rho^2+\eta_T^2)}-\kappa\rho
\ .
\ee
Hence one cannot really associate each component of $\vec\eta$ to a 
particular contribution to the response. However, by removing the transverse 
nucleonic motion (setting $\eta_T=0$), the expressions (\ref{UL}) 
and (\ref{ULPV})
with ${\cal D}^L=0$ are exactly recovered.

Turning now to the the functions ${\cal D}^{L,T,T'}$,
we first notice that ${\cal D}^L$ and ${\cal D}^T$ 
arise from the second term on the right-hand side of (\ref{fmunu}), whereas
${\cal D}^{T'}$ arises from the third one. By explicitly performing the
calculation one arrives at the simple results 
\ba
\label{del}
{\cal D}^L(\kappa,\tau) &=& {\cal D}^T(\kappa,\tau) =
\frac{1}{\epsilon_F-\epsilon_0}\int_{\epsilon_0}^{\epsilon_F} \eta_T^2 
d\epsilon
\nonumber\\ 
&=& 
\frac{\tau}{\kappa^2} \left[ (\lambda\rho+1)^2+(\lambda\rho+1)(1+
\psi^2)\xi_F+\frac{1}{3}(1+\psi^2+\psi^4)\xi_F^2\right] \nonumber\\
&& -(1+\tau\rho^2)
\ea
and
\ba
{\cal D}^{T'}(\kappa,\tau) &=& \frac{1}{\epsilon_F-\epsilon_0}
\int_{\epsilon_0}^{\epsilon_F}
 \left(\sqrt{1+\frac{\eta_T^2}{1+\tau\rho^2}}-1\right) d\epsilon
\nonumber\\
&=&
\frac{1}{\kappa} \sqrt{\frac{\tau}{1+\tau\rho^2}}
\left[1+\xi_F(1+\psi^2)+\lambda\rho\right]-1 \ ,
\label{dela}
\ea
which, when $\rho$=1, reduce to the corresponding quantities in the nucleonic
sector.
These results show that ${\cal D}^L$ and ${\cal D}^T$
just correspond to the mean square value
of the transverse momentum of the nucleon, as already pointed out for the
longitudinal channel \cite{Amo97}, whereas ${\cal D}^{T'}$ is related
to the transverse kinetic energy of the nucleon when $\tau$ is small.
Moreover, not only do the functions ${\cal D}^{L,T,T'}$ {\it start to get}
contributions from
the quadratic terms of the $\eta$ expansion, but actually
${\cal D}^L$ and ${\cal D}^T$, unlike ${\cal D}^{T'}$, are
{\it only} contributed to by the quadratic terms. In order to further  
illustrate the different physical meaning of ${\cal D}^{L(T)}$ and
${\cal D}^{T'}$, we go to the QE ($\Delta$) peak, where $\lambda=\tau \rho$.
Then, from (\ref{del}),(\ref{dela}), it follows that    
\be
\label{delp}
{\cal D}^{L(T)}_P(\tau)= \xi_F + \frac{1}{3 (1 + \tau \rho^2)} \xi_F^2 \,,
\ee
and
\be
{\cal D}^{T'}_P=\frac{\xi_F}{2 (1 + \tau \rho^2)} \,.
\ee
Since $\xi_F$ is quite small ($\xi_F=0.028$ for $^{12}C$), the second term of
(\ref{delp}) is negligible with respect to the first, so that
${\cal D}^{L(T)}_P$ is practically constant and independent of $\tau$
and $\rho$. On the other hand, ${\cal D}^{T'}_P$ vanishes at large $\tau$, and
is smaller at the $\Delta$ peak than at the QE one.  

If the terms associated with ${\cal D}^T$ and ${\cal D}^{T'}$ are neglected,
then the PC transverse and the PV axial responses can be related for 
symmetric nuclei. For the QE process, one gets  
\be
\frac{\tilde{R}^{T'}}{R^T} \simeq \sqrt{\frac{\tau + 1}{\tau}} 
\frac{ G_{M,p}(\tau) \tilde{G}_{A,p}(\tau) + 
G_{M,n}(\tau) \tilde{G}_{A,n}(\tau)}{G^2_{M,p}(\tau) + G^2_{M,n}(\tau)} \,,
\label{r1}
\ee
which can be simplified by neglecting the isoscalar form factors, leading to
\be
\label{r2}
\frac{\tilde{R}^{T'}}{R^T} \simeq \sqrt{\frac{\tau + 1}{\tau}} 
\frac{G_A^{(1)}(\tau)}{G_M^{(1)}(\tau)} \,;
\ee
here $G_{M(A)}^{(1)}= G_{M(A),p} - G_{M(A),n}$ are the isovector magnetic and
axial form factors. The relation (\ref{r2}) corresponds to the
prescription given in
formula (38) of \cite{Bar94}, now proven to be valid up to
next-to-leading order in a nucleon momentum expansion and at all orders in
$\kappa$. Analogously, for the
$N\to\Delta$ transition, we get  
\be
\label{r3}
\frac{\tilde{R}^{T'}}{R^T} \simeq 4 \frac{\mu-1}{\mu+1} \frac{\sqrt{\tau
(\tau \rho^2 + 1)}}{4 \tau + (1-\mu)^2} 
\frac{3  G_{E,\Delta}(\tau) \tilde G_{M,\Delta}^A(\tau) 
+ G_{M,\Delta}(\tau) \tilde G_{E,\Delta}^A(\tau)}{G_{M,\Delta}^2(\tau)+3
G_{E,\Delta}^2(\tau)}\,,
\ee
which, under the assumption of $M1$ dominance of the electromagnetic
$N\to\Delta$ transition, becomes
\be
\label{r4}
\frac{\tilde{R}^{T'}}{R^T} \simeq 4 \frac{\mu-1}{\mu+1} \frac{\sqrt{\tau
(\tau \rho^2 + 1)}}{4 \tau + (1-\mu)^2} \frac{\tilde
G_{E,\Delta}^A(\tau)}{G_{M,\Delta}(\tau)}\,.
\ee
Since ${\cal D}^T$ differs from ${\cal D}^{T'}$ through terms of order
$\eta_F^4$, the same will occur for the two responses; hence, the relationships
(\ref{r1})-(\ref{r4}) are valid to a remarkable accuracy in the RFG model.

Now, as done above for $U$ and ${\cal D}$, it would appear natural also to 
explore the universal factor 
$R_0$ in the light of the $\eta$ expansion.
However, in order to appreciate the importance of a fully relativistic
treatment of the kinematics, we shall rather focus on the difference
\be
R_0 - R_{0,nr}
\ee
between the relativistic 
expression (\ref{R0}) and the  non-relativistic one
\cite{Alb78}, which obtains by assuming non-relativistic 
energy-momentum relations and setting $1/(\epsilon \epsilon') \approx 1$ in 
the energy denominators appearing in (\ref{Wmunu}).
With these approximations one gets
\be
R_{0,nr}= \frac{3 {\cal N}}{8 m_N \kappa \eta_F} (1-\psi_{nr}^2)
\ee
with a non-relativistic scaling variable $\psi_{nr}$ whose square is
given by \cite{Alb90}
\be
\psi^2_{nr}=\left(\frac{\eta_{0,nr}}{\eta_F} \right)^2 
\label{psinr}
\ee
in the Pauli unblocked region.
In the above
\be
\eta_{0,nr} = \frac{2}{\mu-1} \left|\kappa-\sqrt{\mu\left[\kappa^2
-(\mu-1)\lambda\right]}\right|
\ee
represents the minimum momentum a struck nucleon should have to take part
in the process.
In passing it is easily checked that in the limit $\mu=1$ the 
non-relativistic scaling variable 
$\psi_{nr}=[\lambda/\kappa-\kappa]/\eta_F$ of \cite{Alb88} is recovered.

Next, we expand the function
\ba
R_0-R_{0,nr} &=& \frac{3{\cal N}}{4 m_N\kappa\eta_F^3}
\left[\xi_F(1-\psi^2)-\frac{\eta_F^2}{2}(1-\psi_{nr}^2)\right] 
\label{diff1}
\ea
in powers of $\eta_F$. For this purpose we first notice that 
\be
\kappa^2=\tau(1+\tau\rho^2)+2\tau\rho(\lambda-\tau\rho)+{\cal O}(\eta_F^2)\ .
\ee
Then, from (\ref{psi}) and (\ref{psinr}), one gets that
\be
\xi_F\psi^2 = {\cal O}(\eta_F^2) \,,
\ee
whereas
\be
\eta_F^2\psi_{nr}^2 = 4\tau\left[
\frac{\sqrt{1+\tau\rho^2}-\sqrt{\mu\left[1+\tau\rho^2-\rho(\mu-1)\right]
}}{\mu-1}\right]^2
+{\cal O}(\eta_F)\ .
\ee
We thus reach the conclusion that the quantity
\be
R_0-R_{0,nr} = \frac{3{\cal N}}{4 m_N\kappa\eta_F^3}\left\{2\tau\left[
\frac{\sqrt{1+\tau\rho^2}-\sqrt{\mu\left[1+\tau\rho^2-\rho(\mu-1)\right]
}}{\mu-1}
\right]^2
+{\cal O}(\eta_F)\right\}
\label{diff}
\ee
not only does not vanish at leading order in $\eta_F$, 
but can be quite large indeed for high values of $\tau$.
In particular, in the limit 
$\mu\to 1$, (\ref{diff}) yields
\be
R_0-R_{0,nr} = \frac{3{\cal N}}{8 m_N\kappa\eta_F^3}\left[
\frac{\tau^3}{1+\tau}+{\cal O}(\eta_F)\right] \ ,
\ee
from which it follows that $R_{0,nr}$ should {\it not} be used if $\tau$ is 
not small.

We recall, however, that in the nucleonic sector an approximate scaling
variable has been suggested in \cite{Alb90}, namely
\be
\tilde\psi_{nr} = \frac{1}{\eta_F}\left[\frac{\lambda(1+\lambda)}
{\kappa}-\kappa\right]\ ,
\label{psitil}
\ee
which, although displaying the typical non-relativistic structure,
actually incorporates some effects of relativity in the 
$\lambda(\lambda+1)$ term.
The latter, when expanded in the nucleonic momentum, leads to 
\be
\eta_F\tilde\psi_{nr} = \frac{\lambda-\tau}{\sqrt{\tau(1+\tau)}}\ .
\label{etafpsitil}
\ee
By inserting the above into (\ref{diff1}) one obtains a difference of order 
$\eta_F^2$ between the exact and approximate $R_0$. This is another
instance where using the leptonic variables $\kappa$, $\lambda$ and
$\tau$ judiciously leads to improved convergence of expansions in
$\eta_F$.

In summary, in this letter we have re-examined the interplay between
the single-baryon and nuclear many-body content that takes place in
electroweak interactions with nuclei. We have limited our focus to
inclusive responses and have taken the RFG for guidance in attempting
to gain insight into how these two elements enter. In particular, we
have focused on expansions of the single-baryon cross sections in the
one small parameter in the problem that is available in high-energy
studies, namely, the dimensionless ratio $\eta_F=p_F/m_N$. Upon
judiciously making use of the factor $\kappa^2/\tau=|q^2/Q^2|$, which
is greater than unity and embodies some of the moving nucleon content
in the problem, we are able to obtain single-baryon cross sections
whose leading terms contain no additional $\eta_F$-dependence beyond
that implicit in factors such as $\kappa^2/\tau$, and have in 
next-to-leading order only contributions of order $\eta_F^2$, with no
linear terms. Such corrections are usually so small that they
can safely be neglected, providing considerable simplification to
descriptions of inclusive electroweak processes. In contrast, for the
nuclear many-body problem itself, as distinct from the single-baryon
substructure, such non-relativistic expansions must be done with care.


\section*{Acknowledgments}
This work is supported in part by the Bruno Rossi MIT-INFN exchange program, by the U.S. Department 
of Energy under Cooperative Research
Agreement No. DF-FC02-94ER40818 and by Spanish DGYCIT contract No. PB 96-0753. 



\newpage
\section*{Figures}

\end{document}